\documentclass[aps,prb,preprint,showpacs,footnoteinbib]{revtex4}
\usepackage{latexsym}
\usepackage{amssymb}
\usepackage{graphicx,epsfig}
\usepackage{amssymb, amsmath}

\begin{document}
\title{Stable switch action based on quantum interference effect}

\author{Sreemoyee Mukherjee}
\email{sree.mukherjee@bose.res.in}

\author{Ashutosh Yadav}
\email{ashutoshauchem@gmail.com} 

\author{P. Singha Deo}
\email{deo@bose.res.in}
\affiliation{S. N. Bose National Centre for Basic Sciences, Salt Lake, Kolkata 700 098, India}

\date{\today}

\begin{abstract}
\noindent
Although devices working on quantum principles can revolutionize the electronic industry, 
they have not been achieved yet as it is difficult to control their stability. We show that 
one can use evanescent modes to build stable quantum switches. The physical principles that 
make this possible is explained in detail. Demonstrations are given using a multichannel 
Aharonov - Bohm interferometer. We propose a new $S$ matrix for multichannel junctions to 
solve the scattering problem.
\end{abstract}



\maketitle

\section{\bf Introduction}
Advances in electron beam lithography within the last few years have made it possible 
to fabricate nano sized or mesoscopic artificial structures with good control over design 
parameters and probe the quantum transport properties \cite{datta}. These include very narrow 
quasi one-dimensional quantum wires, zero-dimensional electron systems or quantum dots, rings, 
etc., constructed at a semiconductor interface. Typical sizes of these systems vary between 1 to 
10 ${\mu} $m. At very low temperatures (typically mK), the scattering by phonons is significantly 
suppressed, and the phase coherence length can become large compared to the system size. In this 
regime the electron maintains the single particle phase coherence across the entire sample. The 
sample becomes an electron waveguide where the transport properties are solely determined by the 
impurity configuration and the geometry of the conductor and by the principles of quantum 
mechanics \cite{datta}.

Such advances in mesoscopic structures have led to the possibility of new 
quantum semiconductor devices. These active quantum devices rely on quantum 
effects for their operation based on interferometric principles, and are 
quantum analog of well-known optical and microwave devices \cite{datta}. 
The mechanism of switch action by quantum interference is a new idea in electronic 
application. Several potential switching devices have been proposed, wherein one 
controls the relative phase difference between different interfering paths 
(say, in semiconducting loop structures) by applying electrostatic or magnetic fields 
\cite{dat85, dat90, datrpp}. The possibility of achieving transistor action in T-shaped 
structure by varying the effective length of the vertical open ended 
lead has also been explored \cite{sols}. Devices in which electrons carry 
current without being scattered either elastically or inelastically 
(ballistic devices) promise to be much faster and will consume less power 
than the conventional devices. It should also be noted that quantum devices can 
exhibit multifunctional property (e.g., single stage frequency multiplier) wherein 
the functions of an entire circuit within a single element can be performed 
\cite{sub}. They can also lead to tremendous down sizing of electronic devices. 
The conventional transistors operate in a classical diffusive regime 
and are not very sensitive to variations in material parameters such as dimensions or 
the presence of small impurities or non-uniformity in size and shape. These devices 
operate by controlling the carrier density of quasi-particles. However, proposed quantum 
devices are not very robust in the sense that the operational characteristics depend 
very sensitively on material parameters, impurity configuration, shape and size, temperature 
and Fermi energy \cite{land}. For example, incorporation of a single impurity
 in the mesoscopic device can change, non-trivially, the interference of partial 
electron waves propagating through the sample, and hence the electron transmission 
(operational characteristics) across the sample \cite{bcg}. In such devices the actual 
problem of control and reproducibility of operating thresholds 
becomes highly nontrivial. These devices can be exploited if we achieve the 
technology that can reduce or control the phase fluctuations to a small fraction 
of $ {2\pi} $ \cite{deo}. A lot of work has been done in one dimensional 
quantum rings \cite{deo, mol, pet, pet1, pet2, pet3, vasi, kal, jos, ben2}. 
However, the experimental rings are always in two dimension or in three 
dimension. Such systems have not received much theoretical attention because 
multichannel junctions are very difficult to account for theoretically. Earlier 
models either do not account for channel mixing or do not allow 
the inclusion of evanescent modes.

\section{\bf Theoretical Analysis}

\begin{figure}[H]
\centering
\includegraphics[height=6cm,width=8cm]{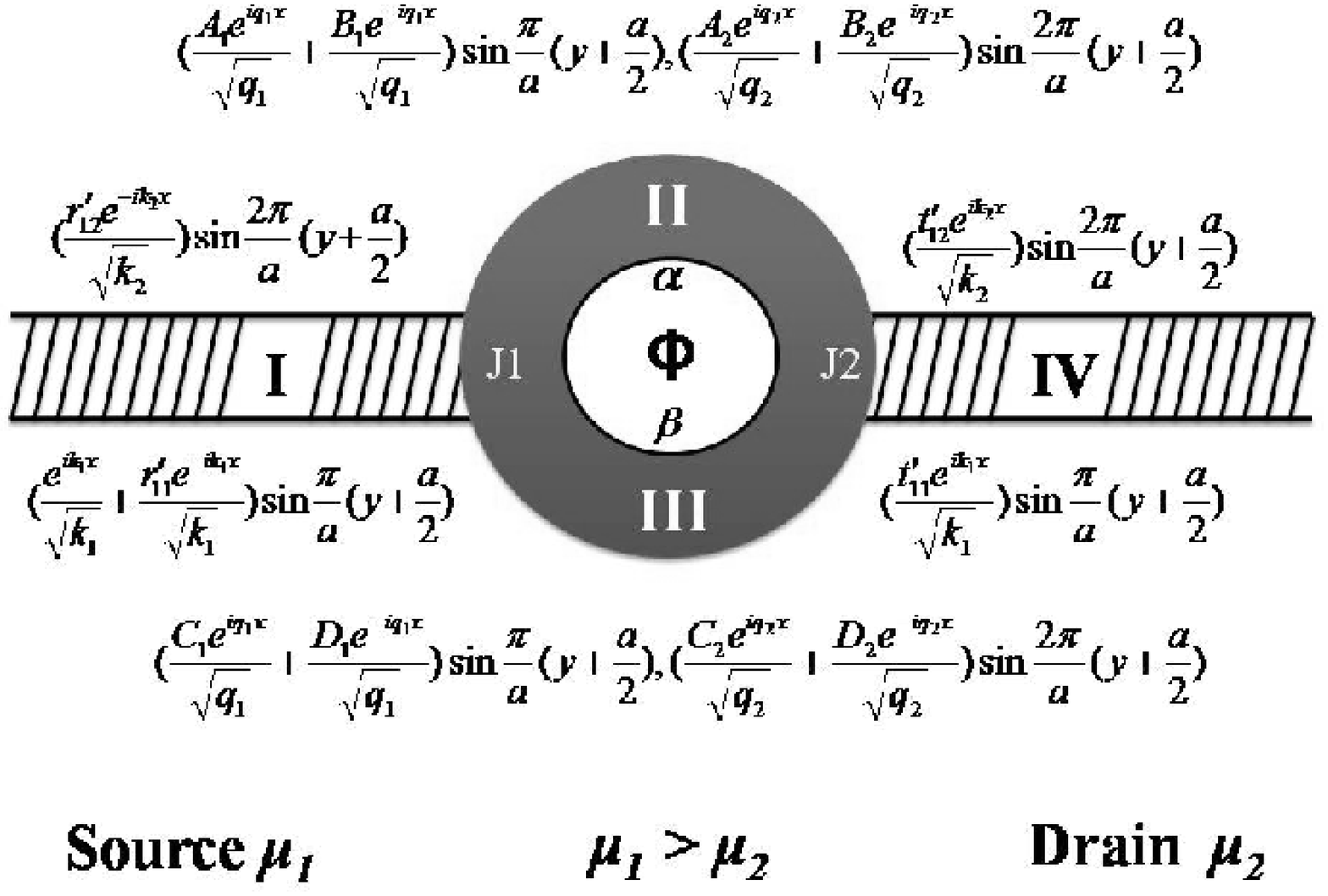}
\caption{\small A finite thickness quantum ring of width $ a $ made up of normal 
metal or semiconductor is indicated by the grey region. On either sides the quantum 
ring is attached with quantum wires made up of normal metal or semiconductor. It is 
indicated by the stripped regions. On the left of the above system there is the source 
reservoir whose chemical potential is $\mu_{1}$ and on the right there is the drain 
reservoir whose chemical potential is $\mu_{2}$.  A potential difference $(\mu_{1}-\mu_{2}) $ 
between the source reservoir and the drain reservoir drives a transport current. The wave 
functions of the electron in different regions is shown in the figure at their respective 
places. Different regions are marked as I, II, III and IV. The ring is pierced by an 
Aharonov - Bohm flux $ \phi $. $ \alpha $ is the Aharonov- Bohm phase an electron picks 
up in region II and $ \beta $ is that in region III. $ J1 $ is the junction where the 
regions I, II and III meets and $ J2 $ is the junction where the regions II, III and IV meets. }
\label{fig1}
\end{figure}
Fig. \ref{fig1} represents the schematic diagram of a finite thickness mesoscopic 
quantum ring under consideration. It is made up of normal metal or semiconductor and 
electronic transport in such systems can be well described by an effective mass theory 
\cite{datta}. Incident electrons coming from the source reservoir on the left (say), gets 
scattered by the ring. Division of wave front occurs at junction $ J1 $; a partial wave 
propagates along the upper arm of the ring and another partial wave propagates along the 
lower arm of the ring. These two partial waves recombine and give a transmittance that 
bears the signature of interference between the two partial waves along the two arms of the 
ring. This interference can be modified by an Aharonov - Bohm flux through the center of the ring. 
The description of the figure is given in further detail in the figure caption. 

The Schr$\ddot{o}$dinger equation for a qausi one dimensional wire is (the third 
degree of freedom, i.e., z-direction, is usually frozen by creating a strong quantization \cite{datta})  
\begin{equation}
-{{{\hbar}^2} \over 2m^*}({{\partial}^2{\psi} \over {{\partial}x^2}}+{{\partial}^2{\psi} \over {{\partial}y^2}}) + V(x,y){\psi}(x,y) = E{\psi}(x,y)
\label{eq1}
\end{equation} 
Here the $ x $ coordinate is along the wire, y coordinate is perpendicular to it, $ m^* $ 
is the electron effective mass and E is the electron energy. In region I and IV (see Fig. 1) 
we have only the confinement potential. That is $$ V(x,y) =  V(y) .$$ Whereas in region II and 
III apart from the confinement potential we take a constant potential $ V_0 $ that can be used 
to excite evanescent modes inside the ring. That is $$ V(x,y) = V(y) + V_0. $$ Without any loss
 of generality we take $ V(y) $ to be an infinite square well potential. That is\\
$$ V(y)=0\quad\text{for} -a/2\leq y\leq a/2 $$   
and
\begin{equation}
V(y)= \infty\quad\text{for} |y|> a/2
\label{eq2}
\end{equation}
The width of the quantum wire is $ a $. The wave functions in the ring can be obtained by 
solving Eq. (\ref{eq1}) where we assume the ring to be so large compared to the de Broglie 
wave length that its curvature can be neglected \cite{she}. The length of the ring is $ L= l_1 + l_2 $, 
where $ l_1 $ is the length of the upper arm and $ l_2 $ is the length of the lower arm. 
The magnetic field appears just as a phase of $ {\psi}(x,y) $ and will be accounted for while 
applying the boundary conditions \cite{che}. 
In regions I and IV Eq. (\ref{eq1}) can be separated as
\begin{equation} 
\psi(x,y) = \phi(x)\xi(y) 
\label{eq3}
\end{equation}
to give
\begin{equation}
-{{{\hbar}^2} \over 2m^*}{{{\partial}^2{\phi}(x)}\over {{\partial}x^2}} = {{{\hbar^2}k^2}\over 2m^*}{\phi}(x) 
\label{eq4} 
\end{equation}
and
\begin{equation}
-{{{\hbar}^2} \over 2m^*}{{{\partial}^2{\xi}(y)}\over {{\partial}y^2}}+V(y){\xi}(y) = \varepsilon{\xi}(y) 
\label{eq5}
\end{equation}
Since $ V(y)$ is a square well potential of width $ a $, Eq. (\ref{eq5}) gives 
\begin{equation}
{\xi}_n(y)={\sin}{{n\pi}\over a}({a\over 2} +y)
\label{eq6}
\end{equation} and 
\begin{equation}
\varepsilon_n = {{n^2{\pi}^2{\hbar}^2}\over{2m^*a^2}}
\label{eq7} 
\end{equation}
Eq. (\ref{eq4}) has solution of the form 
\begin{equation}
\phi_n(x) = e^{{\pm}ik_nx}$$ with $$ k_n=\sqrt{\frac{2m^*E}{\hbar^2}-\frac{n^2\pi^2}{a^2}} 
\label{eq8}
\end{equation}
or 
\begin{equation}
E= \varepsilon_n + {{{\hbar^2}k_n^2}\over 2m^*}
\label{eq9}
\end{equation}
So wave functions in different regions I and IV can be written as
\begin{equation}
{\psi_I}^{(1)} = ({e^{ik_1x}\over\sqrt k_1}+{r^{\prime}_{11}e^{-ik_1x}\over\sqrt k_1})\sin{{\pi}\over a}(y+{a\over 2}) 
\label{eq10}
\end{equation}
\begin{equation}
{\psi_I}^{(2)} = ({r^{\prime}_{12}e^{-ik_2x}\over\sqrt k_2})\sin{2{\pi}\over a}(y+{a\over 2}) 
\label{eq11}
\end{equation}
\begin{equation}
{\psi_{IV}}^{(1)} = ({t^{\prime}_{11}e^{ik_1x}\over\sqrt k_1})\sin{{\pi}\over a}(y+{a\over 2})
\label{eq12}
\end{equation}
\begin{equation}
{\psi_{IV}}^{(2)} = ({t^{\prime}_{12}e^{ik_2x}\over\sqrt k_2})\sin{2{\pi}\over a}(y+{a\over 2})
\label{eq13}
\end{equation}
$ {\psi_I}^{(1)} $ is the wave function of region I in channel $ n = 1 $ and so on.
From Eq. (\ref{eq8}), in the first mode
\begin{equation}
k_1=\sqrt{\frac{2m^*E}{\hbar^2}-\frac{\pi^2}{a^2}}
\label{eq14}
\end{equation}
is the propagating wave vector and in the second mode 
\begin{equation}
k_2=\sqrt{\frac{2m^*E}{\hbar^2}-\frac{4 \pi^2}{a^2}}
\label{eq15}
\end{equation}
is the propagating wave vector. For 
\begin{equation}
\frac{4 \pi^2}{a^2} < E < \frac{9 \pi^2}{a^2}
\label{eq16}
\end{equation}
both $ k_1 $ and $ k_2 $ are real as can be seen from Eq. (\ref{eq14}) and Eq. (\ref{eq15}) 
and $ k_n $ for $ n>2 $ are imaginary as can be seen from Eq. (\ref{eq8}) implying that 
there are two propagating channels. In the leads we can not have evanescent modes \cite{wee, wha}. 
Now for the region II and region III the potential is $ V(x,y)= V(y) + V_0 $. Wave functions in 
these regions can be similarly written as
\begin{equation}
{\psi_{II}}^{(1)} = ({A_1e^{iq_1x}\over\sqrt q_1}+{B_1e^{-iq_1x}\over\sqrt q_1})\sin{{\pi}\over a}(y+{a\over 2})
\label{eq17}
\end{equation}
\begin{equation}
{\psi_{II}}^{(2)} = ({A_2e^{iq_2x}\over\sqrt q_2}+{B_2e^{-iq_2x}\over\sqrt q_2})\sin{2{\pi}\over a}(y+{a\over 2})
\label{eq18}
\end{equation}
\begin{equation}
{\psi_{III}}^{(1)} = ({C_1e^{iq_1x}\over\sqrt q_1}+{D_1e^{-iq_1x}\over\sqrt q_1})\sin{{\pi}\over a}(y+{a\over 2})
\label{eq19}
\end{equation}
\begin{equation}
{\psi_{III}}^{(2)} = ({C_2e^{iq_2x}\over\sqrt q_2}+{D_2e^{-iq_2x}\over\sqrt q_2})\sin{2{\pi}\over a}(y+{a\over 2})
\label{eq20}
\end{equation}
In these regions the energy can be similarly written as
$$ E - V_0 = {{{\hbar}^2q_n^2}\over 2m^*} + {{n^2{\pi}^2{\hbar}^2}\over{2m^*a^2}} $$
or
$$ q_n = \sqrt{{2m^*(E - V_0)\over {\hbar}^2} - {n^2{\pi}^2\over {a^2}}} $$
Hence, in the first mode
\begin{equation}
q_1 = \sqrt{{2m^*(E - V_0)\over {\hbar}^2} - {{\pi}^2\over {a^2}}}
\label{eq21}
\end{equation}
is the wave vector and in the second mode 
\begin{equation}
q_2 = \sqrt{{2m^*(E - V_0)\over{\hbar}^2} - {{4\pi^2}\over{a^2}}}
\label{eq22}
\end{equation}
is the wave vector. Depending on the choice of energy $ E $ and potential $ V_0 $, $ q_1 $ 
and $ q_2 $ can be real (propagating mode) as well as imaginary (evanescent mode). 
Such evanescent states can always be excited in the internal regions of the system but not in leads \cite{wee, wha}.

{\large\it $S$ - matrix for the Junction}

\begin{figure}[H]
\centering
\includegraphics[height=6truecm,width=8truecm]{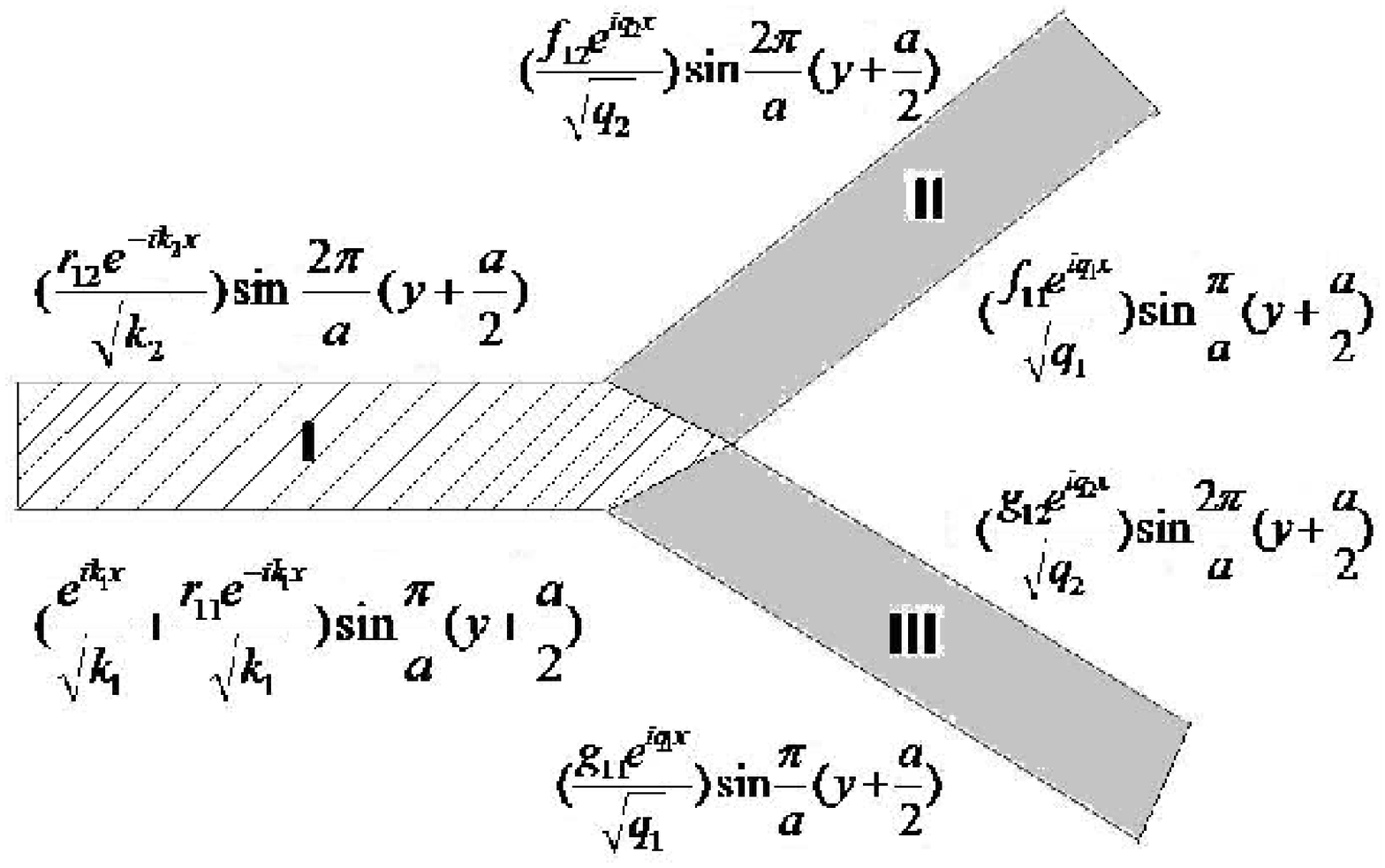}
\caption{\small A three-legged two-channel junction that exists at $ J1 $ 
and $ J2 $ of Fig. \ref{fig1}}
\label{fig2}
\end{figure}

Earlier works have proposed junction $S$ matrix for solving the scattering problem of a 
ring \cite{mol, pet, pet1, pet2, pet3, vasi, kal, jos, but, pin} where the following 
conditions are satisfied at the junction:
(a) conservation of current,
(b) continuity of wave function and
(c) unitarity of $S$ matrix. However, earlier models do not account for channel mixing 
and also do not allow us to include evanescent modes. We give below a simple way to obtain 
an $S$ matrix for a 3 legged two channel junction shown in Fig. \ref{fig2},
that satisfy the three conditions stated above. For our junction $S$ matrix channel 
mixing occurs and evanescent modes can also be accounted for. The approach can be 
generalized to any numbers of channels.

Fig. \ref{fig2} represents schematic diagram of a 3 legged scatterer that we find 
at $ J1 $ or $ J2 $ of Fig. \ref{fig1}. All the legs or leads are made up of normal metal 
or semiconductor. Incident electrons coming from the left of the lead labeled I, gets 
scattered at the junction where the three leads meet. Division of wave front occurs and 
partial waves propagate along the upper arm labeled II and along the lower arm labeled III. 
The wave functions of the electron in different regions are again obtained from Eq. (\ref{eq1}) 
and shown in their respective places for two propagating channels in each lead. Potential in 
region I (stripped region) is zero whereas potential in regions II and III (shaded region) is $ V_0 $.

Taking the clue from reflection and transmission amplitudes for a one dimensional step 
potential \cite{merz} we can write for the different reflection amplitudes $ r_{mn} $ 
and transmission amplitudes $ f_{mn} $ and $ g_{mn} $ shown in Fig. \ref{fig2} as 
$$ r_{11} = ({{k_1 - k_2 - 2q_1 - 2q_2}\over {k_1 + k_2 + 2q_1 + 2q_2}}) $$
$$ r_{12} = ({{2{\sqrt{{k_1}{k_2}}}}\over {k_1 + k_2 + 2q_1 + 2q_2}}) $$
$$ f_{11} = g_{11} = ({{2{\sqrt{{k_1}{q_1}}}}\over {k_1 + k_2 + 2q_1 + 2q_2}}) $$
$$ f_{12} = g_{12} = ({{2{\sqrt{{k_1}{q_2}}}}\over {k_1 + k_2 + 2q_1 + 2q_2}}) $$
$$ r_{22} = ({{k_2-k_1-2q_1 - 2q_2} \over {k_1 + k_2 + 2q_1 + 2q_2}}) $$
$$ r_{21} = ({{2{\sqrt{{k_1}{k_2}}}}\over {k_1 + k_2 + 2q_1 + 2q_2}}) $$
$$ f_{21} = g_{21} = ({{2{\sqrt{{k_2}{q_1}}}}\over {k_1 + k_2 + 2q_1 + 2q_2}}) $$
$$ f_{22} = g_{22} = ({{2{\sqrt{{k_2}{q_2}}}}\over {k_1 + k_2 + 2q_1 + 2q_2}}) $$
$S$ matrix for the junction $ S_j $ is therefore
\begin{equation}
S_j = 
\left[ {\begin{tabular}{cccccc}
  $r_{11}$ & $r_{12}$ & $f_{11}$ & $f_{12}$ & $g_{11}$ & $g_{12}$\\
  $r_{21}$ & $r_{22}$ & $f_{21}$ & $f_{22}$ & $g_{21}$ & $g_{22}$\\
  $f_{11}$ & $f_{12}$ & $r_{11}$ & $r_{12}$ & $g_{11}$ & $g_{12}$\\
  $f_{21}$ & $f_{22}$ & $r_{21}$ & $r_{22}$ & $g_{21}$ & $g_{22}$\\
  $g_{11}$ & $g_{12}$ & $g_{11}$ & $g_{12}$ & $r_{11}$ & $r_{12}$\\
  $g_{21}$ & $g_{22}$ & $g_{21}$ & $g_{22}$ & $r_{21}$ & $r_{22}$\\
\end{tabular}  }
\right]
\label{eq23}
\end{equation}
One can check that the following conditions of unitarity are satisfied
\begin{equation}
|r_{11}|^2 + {|r_{12}|^2} + {|f_{11}|^2} + {|f_{12}|^2} + {|g_{11}|^2} + {|g_{12}|^2} = 1
\label{eq24}
\end{equation}
\begin{equation}
|r_{22}|^2 + {|r_{21}|^2} + {|f_{21}|^2} + {|f_{22}|^2} + {|g_{21}|^2} + {|g_{22}|^2} = 1
\label{eq25}
\end{equation}
The ring wave functions and the lead wave functions at the junction $J1$ of Fig. \ref{fig1}
 can be matched as
\begin{equation}
\left( {\begin{tabular}{c}
  ${r'_{11}\over\sqrt {k_1}}$\\
  ${r'_{12}\over\sqrt {k_2} }$\\
  ${A_1\over\sqrt {q_1}}$\\
  ${A_2\over\sqrt {q_2}}$\\
  ${D_1e^{-iq_1l_2}\over\sqrt {q_1}}$\\
  ${D_2e^{-iq_2l_2}\over\sqrt {q_2}}$\\
\end{tabular}  }
\right)
= S_j
\left( {\begin{tabular}{c}
  $1\over\sqrt {k_1}$\\
  $0$\\
  $B_1e^{-i\alpha}\over\sqrt {q_1}$\\
  $B_2e^{-i\alpha}\over\sqrt {q_2}$\\
  $C_1e^{iq_1l_2+i\beta}\over\sqrt {q_1}$\\
  $C_2e^{iq_2l_2+i\beta}\over\sqrt {q_2}$\\
\end{tabular}  }
\right)
\label{eq26}
\end{equation}
Similarly one can match the wavefunctions at the junction $J2$ to give a set of equations given below
\begin{equation}
1+r'_{11}-\sqrt{k_1\over{q_1}}A_1-\sqrt{k_1\over{q_1}}B_1e^{-i\alpha} = 0 
\label{eq28}
\end{equation}
\begin{equation}
r'_{12}-\sqrt{k_2\over{q_2}}A_2-\sqrt{k_2\over{q_2}}B_2e^{-i\alpha} = 0 
\label{eq29}
\end{equation}
\begin{equation}
1+r'_{11}-\sqrt{k_1\over {k_2}}r'_{12} = 0
\label{eq30}
\end{equation}
\begin{equation}
1+r'_{11}-\sqrt{k_1\over{q_1}}C_1e^{iq_1l_2+i\beta}-\sqrt{k_1\over{q_1}}D_1e^{-iq_1l_2} = 0
\label{eq31}
\end{equation}
\begin{equation}
r'_{12}-\sqrt{k_2\over{q_2}}C_2e^{iq_2l_2+i\beta}-\sqrt{k_2\over{q_2}}D_2e^{-iq_2l_2} = 0
\label{eq32}
\end{equation}
$$ik_1-ik_1r'_{11}-iq_1\sqrt{k_1\over{q_1}}A_1+iq_1\sqrt{k_1\over{q_1}}B_1e^{-i\alpha}-iq_2\sqrt{k_1\over{q_2}}A_2+$$
$$ iq_2\sqrt{k_1\over{q_2}}B_2e^{-i\alpha}-iq_1\sqrt{k_1\over{q_1}}C_1e^{iq_1l_2+i\beta}-iq_1\sqrt{k_1\over{q_1}}D_1e^{-iq_1l_2}+$$
\begin{equation}
iq_2\sqrt{k_1\over{q_2}}C_2e^{iq_2l_2+i\beta}-iq_2\sqrt{k_1\over{q_2}}D_2e^{-iq_2l_2}-ik_2\sqrt{k_1\over{k_2}}r'_{12} = 0 
\label{eq33}
\end{equation}
\begin{equation}
t^\prime_{11}-\sqrt{k_1\over{q_1}}A_1e^{iq_1l_1+i\alpha}-\sqrt{k_1\over{q_1}}B_1e^{-iq_1l_1} = 0
\label{eq34}
\end{equation}
\begin{equation}
t'_{12}-\sqrt{k_2\over{q_2}}A_2e^{iq_2l_1+i\alpha}-\sqrt{k_2\over{q_2}}B_2e^{-iq_2l_1} = 0 
\label{eq35}
\end{equation}
\begin{equation}
t'_{11}-\sqrt{k_1\over{k_2}}t'_{12} = 0
\label{eq36}
\end{equation}
\begin{equation}
t'_{11}-\sqrt{k_1\over{q_1}}C_1-\sqrt{k_1\over{q_1}}D_1e^{-i\beta} = 0
\label{eq37}
\end{equation}
\begin{equation}
t'_{12}-\sqrt{k_2\over{q_2}}C_2-\sqrt{k_2\over{q_2}}D_2e^{-i\beta} = 0
\label{eq38}
\end{equation}
$$iq_1\sqrt{k_1\over{q_1}}A_1e^{iq_1l_1+i\alpha}-iq_1\sqrt{k_1\over{q_1}}B_1e^{-iq_1l_1}+iq_2\sqrt{k_1\over{q_2}}A_2e^{iq_2l_1+i\alpha}-$$
$$ iq_2\sqrt{k_1\over{q_2}}B_2e^{-iq_2l_1}-iq_1\sqrt{k_1\over{q_1}}C_1+iq_1\sqrt{k_1\over{q_1}}D_1e^{i\beta}-iq_2\sqrt{k_1\over{q_2}}C_2+$$
\begin{equation}
iq_2\sqrt{k_1\over{q_2}}D_2e^{-i\beta}-ik_1t'_{11}-ik_2\sqrt{k_1\over{k_2}}t'_{12} = 0 
\label{eq39}
\end{equation}
Solving them we can find the $ S $ matrix elements $ r'_{11} $, $ r'_{12} $, $ t'_{11} $ 
and $ t'_{12} $ for the Aharonov - Bohm ring.
\section{ Results and Discussions}

Here we are considering two channel Aharonov - Bohm ring that are characterized
 by four transmission amplitudes $ t'_{11} $ , $ t'_{12} $, $ t'_{21} $ and 
$ t'_{22} $ 
and four reflection amplitudes $ r'_{11} $, $ r'_{12} $, $ r'_{21} $ and 
$ r'_{22} $.
 Landauer's formula gives the two probe conductance $ G $ as
\begin{equation}
G= \frac{2e^2}{h}\sum_{i,j}|t'_{ij}|^2.
\label{eq40}
\end{equation}
The transmission amplitude from mode $ j $ to mode $ i $ is $ t'_{ij} $. $ G $ is a strongly 
oscillating function of $ \phi/\phi_0 $ implying we can use flux to drive the system from a 
conducting state to an insulating state that can be identified with 1 and 0 of a switch as will 
be exemplified. Such a switch will therefore be working entirely on quantum mechanical 
principles. Such devices if achieved will be a major technological break through. 
First of all, it will transcend Moore's law by leaps and bounds to result in extremely 
small devices. Secondly, such devices will consume very little power and will solve the 
problem of present day computers dissipating a lot of energy and getting heated up. 
Other advantages are mentioned in introduction. However, such devices have not been 
achieved so far because switches based on quantum interference principles as the one we 
are discussing here in our work are not stable \cite{land}. Small changes in temperature 
or incorporation of a single impurity can drastically change the operational 
characteristics of the switch. One can understand this in terms of the fact 
that impurity cause additional reflections or temperature increases Fermi 
energy and hence wavelength and therefore imply changes in path lengths 
in an interference set up. We demonstrate below how changes in path lengths 
can drastically alter the operational characteristics of an 
Aharonov - Bohm ring. Finally we will show that there is a solution to the problem.

\begin{figure}[H]
\centering
\includegraphics[height=6truecm,width=8truecm]{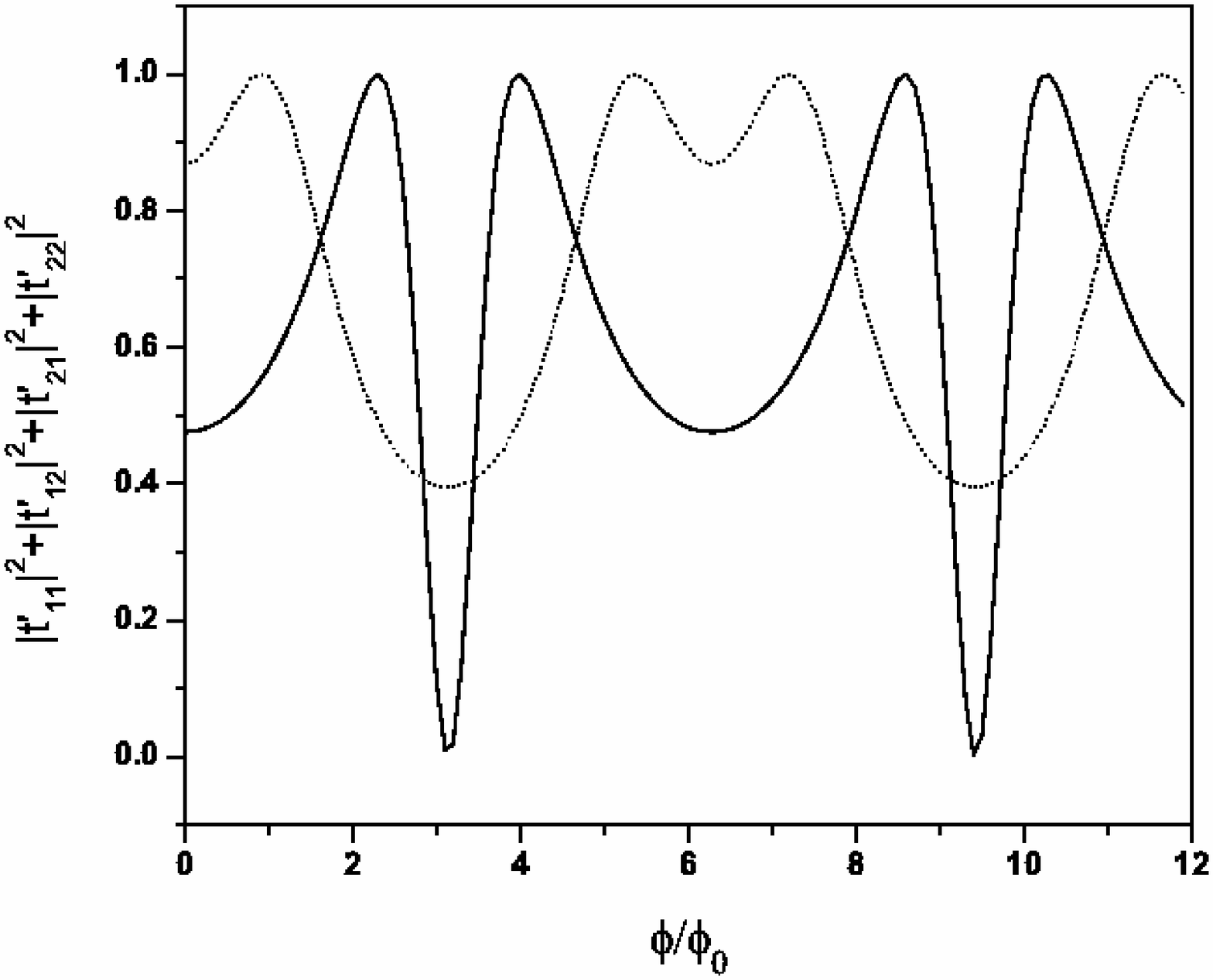}
\caption{ The figure shows a plot of $\sum_{i,j}|t'_{ij}|^2$ as a function of $ \phi/\phi_0 $. 
Here $ \phi_0 = hc/e $. The incoming electrons have 
energy $ E = {55{\hbar}^2\over m^*a^2} $ and the constant potential 
$ V_0 $ of the ring is $ 0 $. We are considering in this case two propagating 
modes. The solid line is $ \sum_{i,j}|t'_{ij}|^2$ for $l_1/a=5$, $l_2/a=5$ 
and the dotted line is $\sum_{i,j}|t'_{ij}|^2 $ for $l_1/a=4$, $ l_2/a=6$.}
\label{fig3}
\end{figure}

\begin{figure}[H]
\centering
\includegraphics[height=6truecm,width=8truecm]{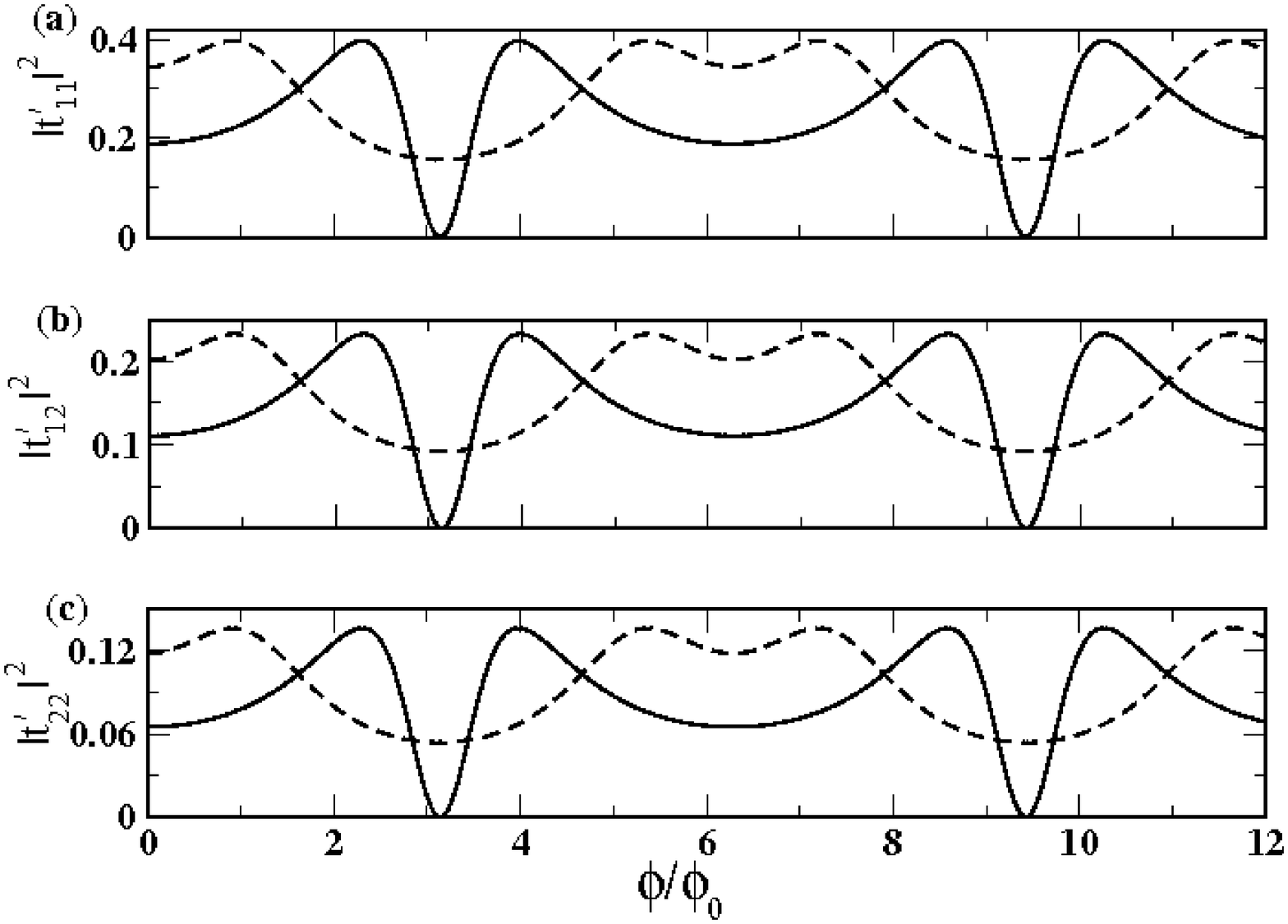}
\caption{\small We use same parameters as in Fig. \ref{fig3} and plot individual scattering
cross sections. (a) shows a plot of $|t'_{11}|^2$ as a function of $ \phi/\phi_0 $. 
(b) shows a plot of $|t'_{12}|^2$ as a function of $ \phi/\phi_0 $. $|t'_{21}|^2$ as 
a function of $ \phi/\phi_0 $ is identical to $|t'_{12}|^2$ as a function of $ \phi/\phi_0 $ 
due to Onsager reciprocality relation, so $|t'_{21}|^2$  
is not shown. (c) shows a plot of $|t'_{22}|^2$ as a function of $ \phi/\phi_0 $. 
Here $ \phi_0 = hc/e $. The solid lines are for $l_1/a=5$, $l_2/a=5$ and the dashed 
lines are for $l_1/a=4$, $ l_2/a=6$.} 
\label{fig4}
\end{figure}

In Fig. \ref{fig3}, we show $ {G\over 2e^2/ h} $ for a two channel Aharonov-Bohm ring with 
$l_1/a=5$, $l_2/a=5$ (solid line) and with $l_1/a=4$, $ l_2/a=6$ (dotted line). 
We choose incident energy in the range given by Eq. (\ref{eq16}) and so we are considering a 
two channel scattering problem. We take the potential inside the ring $ V_0 $ to be $ 0 $ 
implying that both channels are propagating 
inside the ring. Solid line shows two conductance minima, one is shallow at flux $ \phi/\phi_0 = 0 $ 
(approx) and another is deep at flux  
$ \phi/\phi_0 = 3.1 $ (approx) and it also shows one conductance maximum at flux  
$ \phi/\phi_0 = 2.4 $ (approx). Dotted line also shows 
two conductance minima, one is shallow at flux $ \phi/\phi_0 = 0 $ (approx) 
and another is deep at flux  $ \phi/\phi_0 = 3.1 $ (approx)
 and it shows one conductance maximum at flux  $ \phi/\phi_0 = 1.0 $ (approx). 
We can assign the conductance minima as off state and 
conductance maxima as on state of a switch. Fig. \ref{fig3} shows 
that with changing the arm length the minima is not shifting but
 maxima is shifting a lot. The shallow minima for dotted line is so 
shallow that it may not be observed in measurement. 
Much more non-systematic behavior will be shown in subsequent plots. 
Now we will plot the individual $ |t'_{ij}|^2 $ s 
as a function of $ \phi/\phi_0 $ and shown in Fig. \ref{fig4}.

All the plots of individual partial scattering cross sections 
($|t'_{11}|^2$, $|t'_{12}|^2 = |t'_{21}|^2 $ and $|t'_{22}|^2$) are qualitatively 
same as the plot of $\sum_{i,j}|t'_{ij}|^2$ as a function of 
$ \phi/\phi_0 $. Peaks are expected to occur at resonance 
\cite{pas} when integral wave numbers fit into the total 
length of the  Aharonov - Bohm ring. However in presence of channel
 mixing the two channels are not independent. Resonance in one 
channel builds up density of states in the other channel and 
so $ |t'_{11}|^2 $, $ |t'_{12}|^2 $, $ |t'_{21}|^2 $ and 
$ |t'_{22}|^2 $ peak at same flux values. Conductance is
 determined by the addition of these individual scattering cross sections.
 Since they are qualitatively same they add up coherently. 
When all these partial scattering cross sections are coherently 
added the difference between the on state and the off state becomes 
100\% for deep minima and it is 50 \% for shallow minima in case of 
solid line in Fig. \ref{fig3}, while it is only 40\% for deep minima 
and 12\% for shallow minima in case of dotted line in Fig. \ref{fig3}. 
Such variations in magnitudes of drops in conductance apart 
from variations in peak positions already discussed indicates 
that it is not so efficient to make stable switches.

\begin{figure}[H]
\centering
\includegraphics[height=6truecm,width=8truecm]{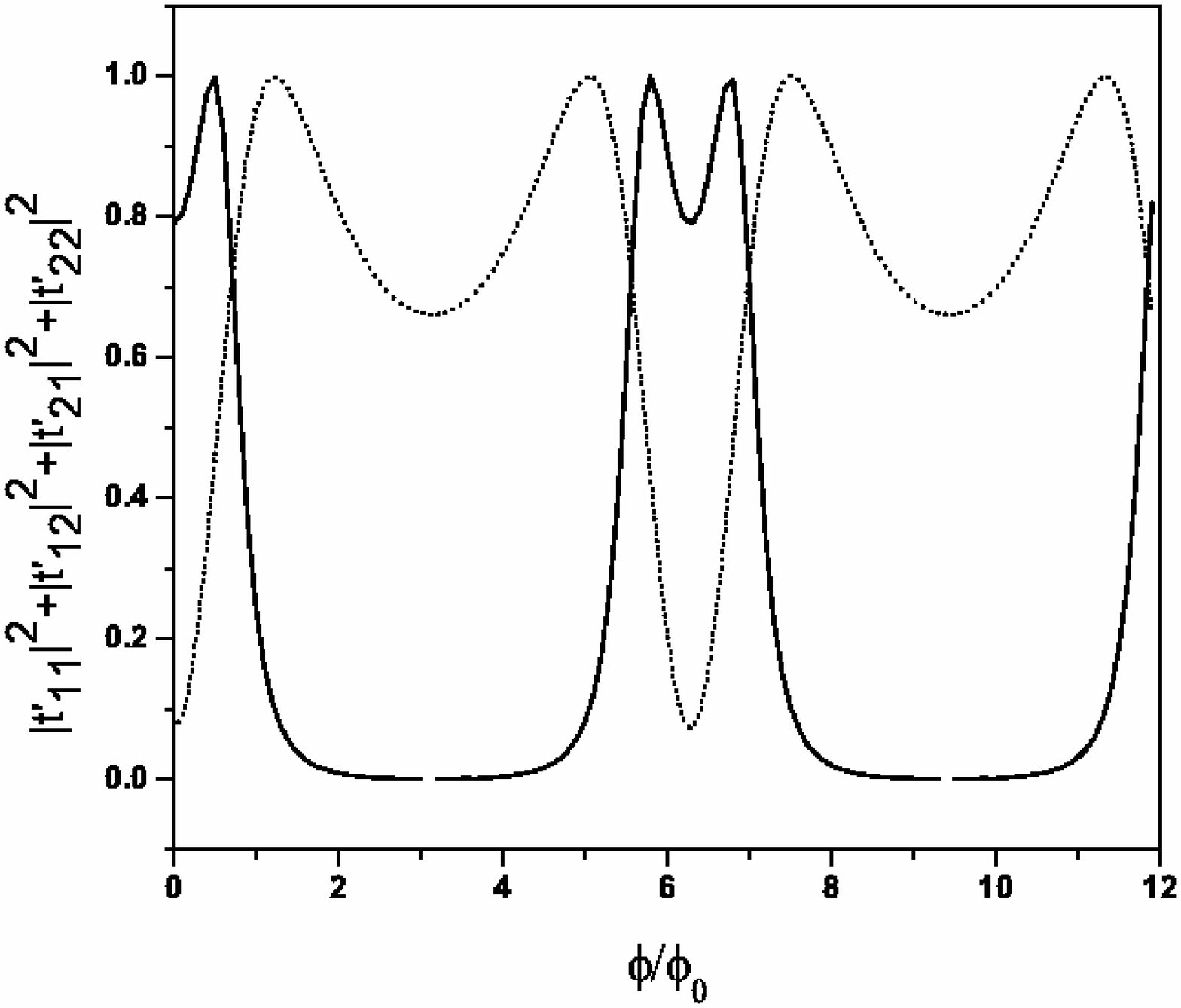}
\caption{\small The figure shows a plot of $\sum_{i,j}|t'_{ij}|^2$ as a function of 
$ \phi/\phi_0 $. Here $ \phi_0 = hc/e $. The incoming electrons have energy $ E = {45{\hbar}^2\over m^*a^2} $, 
the constant potential $ V_0 $ of the ring is such that $ V_0 ={10{\hbar}^2\over em^*a^2} $. 
With this choice $q_1 $ is real and $q_2$ is imaginary. Thus, we are considering in this 
case one propagating mode and one evanescent mode. The solid line is $\sum_{i,j}|t'_{ij}|^2$ 
for $l_1/a=5$, $l_2/a=5$ and the dotted line is $\sum_{i,j}|t'_{i,j}|^2 $ for $l_1/a=3$, $l_2/a=7$.}
\label{fig5}
\end{figure}
\begin{figure}[H]
\centering
\includegraphics[height=6truecm,width=8truecm]{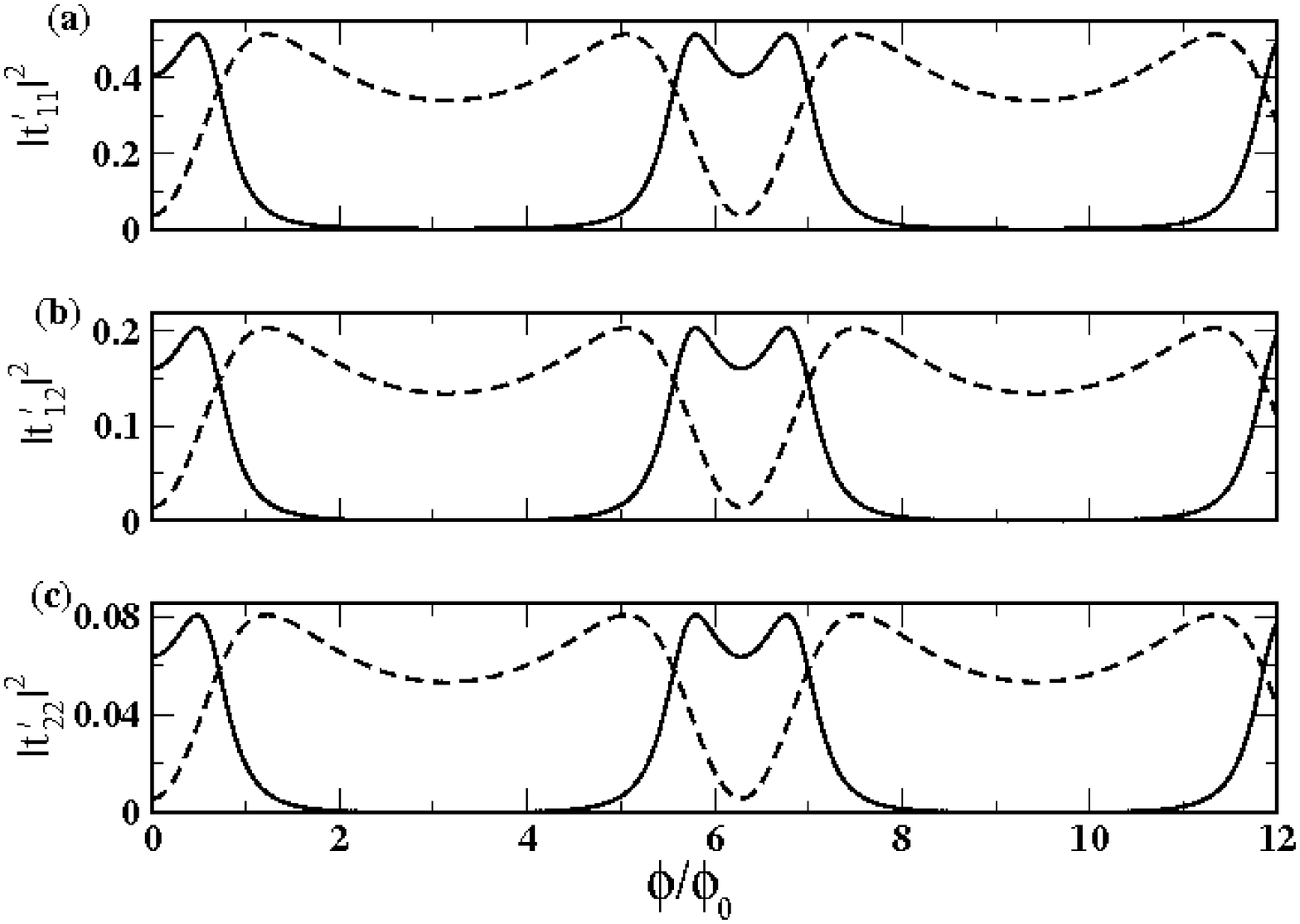}
\caption{\small We use same parameters as in Fig. \ref{fig5} and plot
individual scattering cross sections. (a) shows a plot of $|t'_{11}|^2$ 
as a function of $ \phi/\phi_0 $. 
(b) shows a plot of $|t'_{12}|^2$ as a function of $ \phi/\phi_0 $. $|t'_{21}|^2$ as 
a function of $ \phi/\phi_0 $ is identical to $|t'_{12}|^2$ as a function of $ \phi/\phi_0 $ 
due to Onsager reciprocality relation, so $|t'_{21}|^2$ is not shown. 
(c) shows a plot of $|t'_{22}|^2$ as a function of
 $ \phi/\phi_0 $. Here $ \phi_0 = hc/e $.
The solid lines are for $l_1/a=5$, $l_2/a=5$ 
and the dashed lines are for $l_1/a=3$, $ l_2/a=7$.} 
\label{fig6}
\end{figure}

In Fig. \ref{fig5}, we show $ {G\over 2e^2/ h} $ for a two channel Aharonov-Bohm ring 
with $l_1/a=5$, $l_2/a=5$ (solid line) and with $ l_1/a = 3 $, $ l_2/a = 7 $ (dotted line).
 Again we choose energy in the range given by Eq. (\ref{eq16}). However now we also take a 
non zero electrostatic potential $ V_0 $ inside the ring such that $ q_1 $ is real and $ q_2 $ 
is imaginary (see Eq. (\ref{eq21}) and Eq. (\ref{eq22})). In other words one channel is propagating 
and the other is evanescent. We have checked that such a situation result in just as much diversity as 
that with two propagating modes. Here we demonstrate one particular case and give arguments 
why the behavior is general. Solid line shows two conductance minima and one conductance 
maximum like the solid line in Fig. \ref{fig3}. It has first a shallow minimum and then has a deep 
minimum similar to solid line in Fig. \ref{fig3}. Dotted line shows two conductance minima and 
one conductance maximum like the dotted line in Fig. \ref{fig3}. Here one that was shallow minimum 
in Fig. \ref{fig3} has become a deep minimum and the one that was deep minimum in Fig. \ref{fig3} 
has become shallow minimum. Unlike in Fig. \ref{fig3}, there is wave propagation in only one channel 
 and since the other channel is evanescent, it has no wave propagation. 
Peaks occur for propagating channel when integral wave number 
fits into the total length of the ring \cite{pas}. But here we can see $ |t'_{22}|^2 $ peak at same 
flux values as $ |t'_{11}|^2 $ because again in presence of channel mixing the two modes 
are not independent. Resonance in the propagating channel boost up density of states in the 
evanescent channel and hence the evanescent channel also becomes highly conducting. Thus 
the diversity results from the random behavior of the propagating channel. We show the 
individual $ |t'_{ij}|^2 $s corresponding to Fig. \ref{fig5} in Fig. \ref{fig6}. They are 
again qualitatively same as the curves obtained in Fig. \ref{fig5} which means the individual 
components add up coherently just as it happened for two propagating modes. Conduction along 
the evanescent mode is equally strong and diverse due to the presence of the propagating mode.

\begin{figure}[H]
\centering
\includegraphics[height=6truecm,width=8truecm]{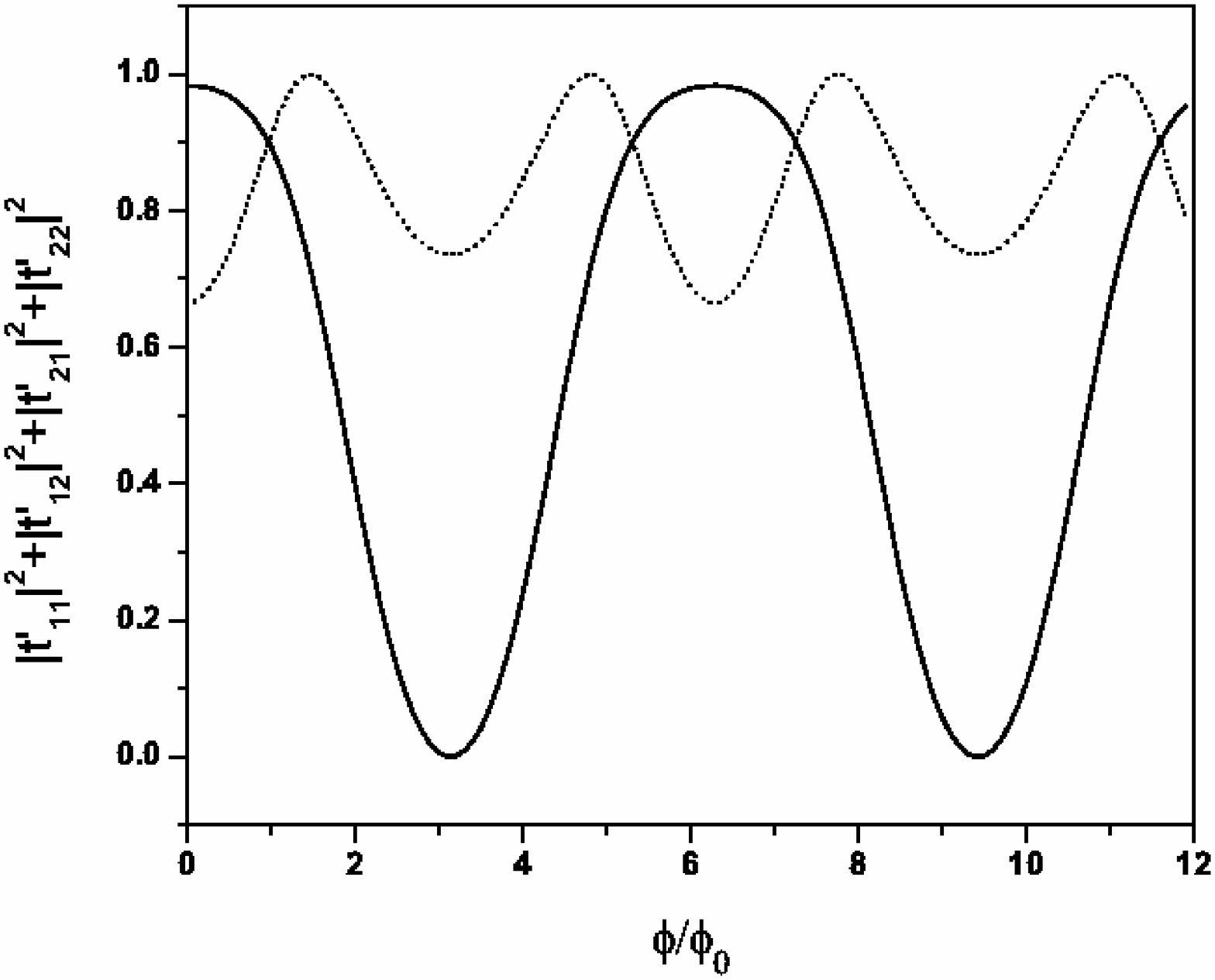}
\caption{\small The figure shows a plot of $\sum_{i,j}|t'_{ij}|^2$ as a function of 
$ \phi/\phi_0 $. Here $ \phi_0 = hc/e $. The incoming electrons have energy 
$ E = {47{\hbar}^2\over m^*a^2} $, the constant potential $ V_0 $ of the ring is 
such that $ V_0 ={10{\hbar}^2\over em^*a^2} $. With this choice $q_1 $ is real and 
$q_2$ is imaginary. Thus, we are considering in this case one propagating mode and 
one evanescent mode. The solid line is $\sum_{i,j}|t'_{ij}|^2$ for $l_1/a=5$, $l_2/a=5$ 
and the dotted line is $\sum_{i,j}|t'_{ij}|^2 $ for $l_1/a=4$, $l_2/a=6$.}
\label{fig7}
\end{figure}
In Fig. \ref{fig7}, we show $ {G\over 2e^2/ h} $ for a two channel Aharonov-Bohm 
ring  with $l_1/a=5$, $l_2/a=5$ (solid line) and with $ l_1/a = 4 $, $ l_2/a = 6$ 
(dotted line). Incident energy and the electrostatic potential are so chosen that 
like in Fig. \ref{fig5} one channel is propagating and another is evanescent. Solid 
line shows one conductance maximum and one conductance minimum instead of two conductance 
minima seen in earlier figures. Only difference is that we have slightly changed the 
energy of the incoming electrons from $ E = {45{\hbar}^2\over m^*a^2} $(Fig. 5) to 
$ E = {47{\hbar}^2\over m^*a^2} $(Fig. 7) keeping $ l_1, l_2 $ same. Dotted line shows 
two conductance minima and one conductance maximum. The drops of two minima are comparable 
which gives the appearance of a $ \phi_0/2 $ periodicity. This is unlike what we saw in 
earlier figures. Here too all the plots of individual partial scattering cross sections 
are qualitatively same as the plot of $\sum_{i,j}|t'_{ij}|^2$ as a function of $ 
\phi/\phi_0 $ and so not shown. 
\begin{figure}[H]
\centering
\includegraphics[height=6truecm,width=8truecm]{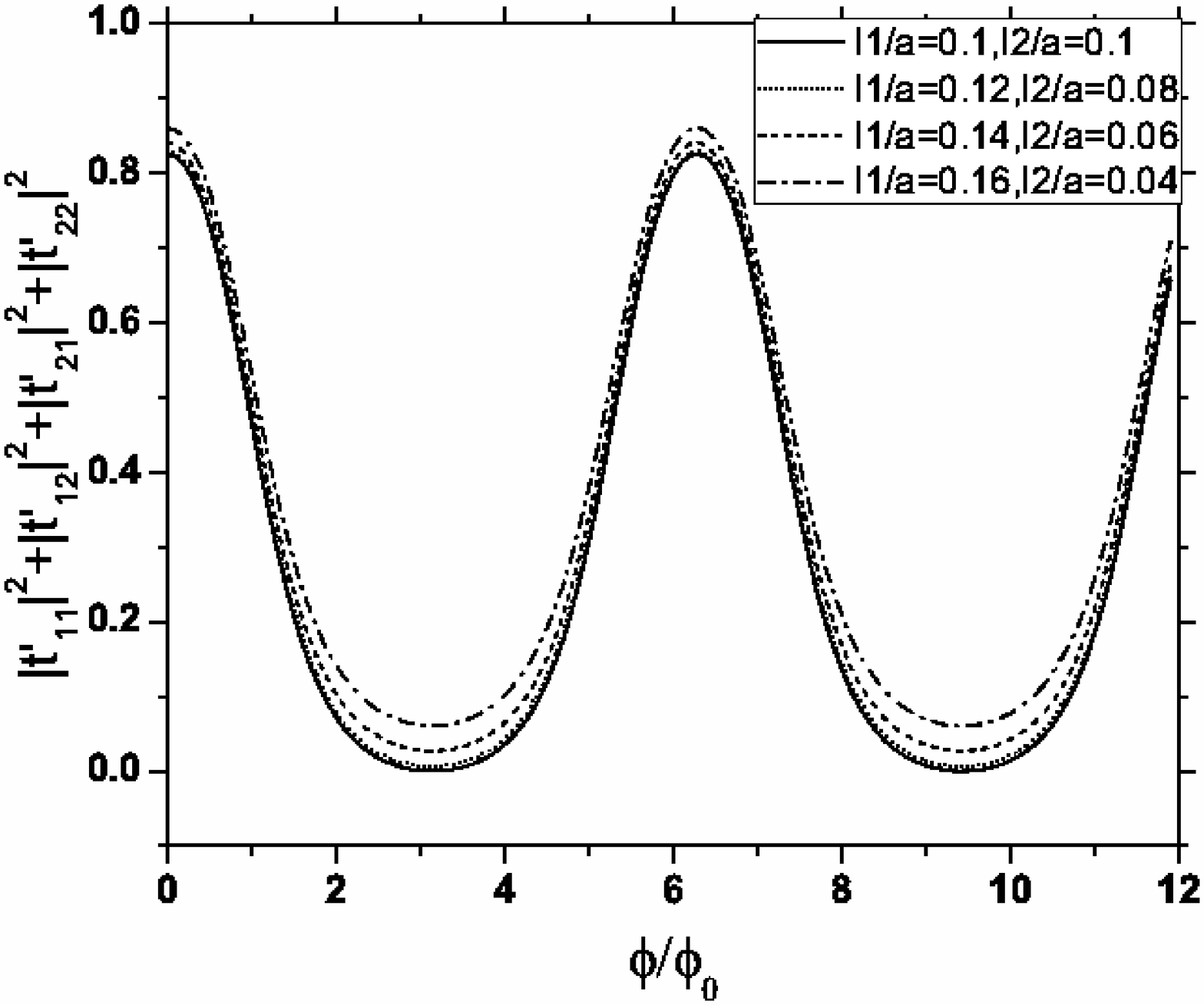}
\caption{\small The figure shows a plot of $\sum_{i,j}|t'_{ij}|^2$ as a function of 
$ \phi/\phi_0 $ for different arm lengths. Here $ \phi_0 = hc/e $. The incoming electrons 
have energy $ E = {49{\hbar}^2\over m^*a^2} $. The constant potential $ V_0 $ of the ring 
is such that $ V_0 ={40{\hbar}^2\over em^*a^2} $. With this choice $q_1 $ and $q_2$ are both 
imaginary. Thus, we are considering in this case two evanescent modes. The exact value of $ l_1 $ and 
$ l_2 $ are given in the figure inset. For all arm length $\sum_{i,j}|t'_{ij}|^2$ as a function 
of $ \phi/\phi_0 $ have the same nature. That means switching action is independent of $ l_1 : l_2 $.}
\label{fig8}
\end{figure}

The random behavior of conductance changes when we make both the modes to be 
evanescent. An electron in an evanescent mode do not acquire phase changes associated
 with propagation. Only phase changes are due to Aharonov - Bohm effect and we find 
that within a period (0 to $ 2\pi $) conductance is maximum at zero flux, then it goes 
through a deep minima and rise again to a maximum value. One can explain this as follows. 
Conductance being a symmetric function of flux (Onsager reciprocality relation), is a 
function of $ (\cos n\phi/\phi_0) $. So it maximizes at $ 0 $ flux and then decreases 
with flux. Periodicity is always $ \phi_0 $ in absence of other competing source of phase 
changes and absence of resonances. This behavior is independent of all parameters as will
 be demonstrated below. However since evanescent modes are not very conducting we have to 
take smaller rings to get same order of magnitude in conductance variations as that of 
propagating modes. But the magnitude can be enhanced by taking a ring that can support many 
evanescent modes.

In Fig. \ref{fig8}, we have taken many choices of arm length and we have seen that 
there is only one conductance maxima at flux $ \phi/\phi_0 = 0 $ (approx) and only 
one conductance minima  at flux $ \phi/\phi_0 = 2.5 $ (approx). Thus, here the behavior 
remains uniform with changing the arm length. Conductance variation from maximum to minimum 
is 82\%. 
\begin{figure}[H]
\centering
\includegraphics[height=6truecm,width=8truecm]{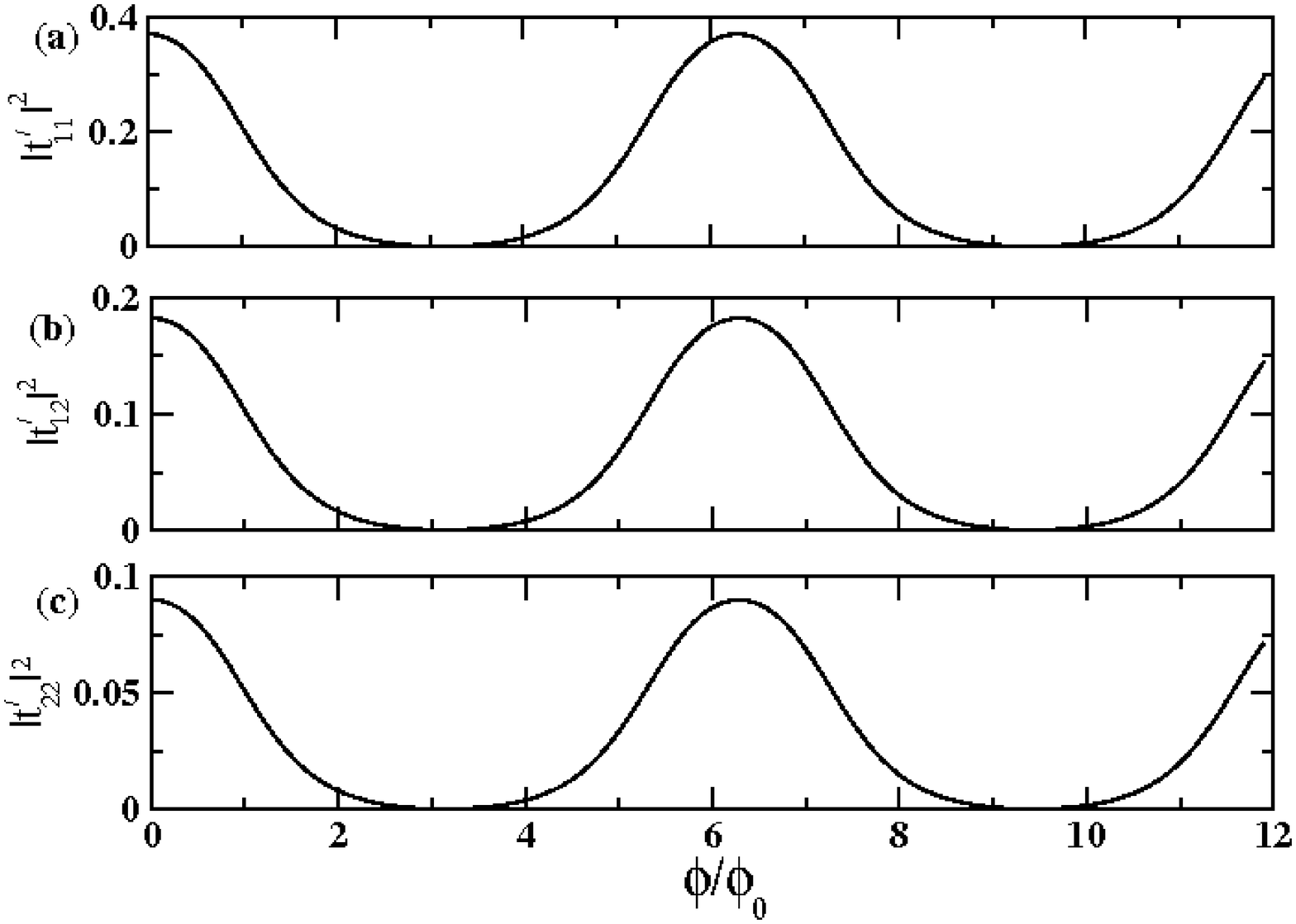}
\caption{\small We use same parameters as in Fig. \ref{fig8} and
plot individual scattering cross sections. 
(a) shows a plot of $|t'_{11}|^2$ as a function of $ \phi/\phi_0 $. 
(b) shows a plot of $|t'_{12}|^2$ as a function of $ \phi/\phi_0 $. $|t'_{21}|^2$ as 
a function of $ \phi/\phi_0 $ is identical to $|t'_{12}|^2$ as a function of $ \phi/\phi_0 $ 
due to Onsager reciprocality relation, so $|t'_{21}|^2$   
is not shown. (c) shows a plot of $|t'_{22}|^2$ as a function of $ \phi/\phi_0 $. 
Here $ \phi_0 = hc/e $. The solid lines are for $l_1/a=0.1$, $l_2/a=0.1$.} 
\label{fig9}
\end{figure}
The individual $ |t'_{ij}|^2 $ s as a function of $ \phi/\phi_0 $ shows similar 
behavior as that of Fig. \ref{fig8} and shown in Fig. \ref{fig9}. 
In case of Fig. 9(a), conductance drops by 38\%, 
in case of Fig. 9(b), conductance drops by 18\% and for Fig. 9(c), conductance drops by 8\%. 
When different channels add up coherently percentage drop of conductance becomes 82\%. By 
using larger and larger number of evanescent channels percentage drop in conductance can 
thus be enhanced and efficiency of switch can be increased. 

It is not always possible to maintain the incidence energy (${Em^*a^2\over{\hbar}^2} $) values 
constant due to statistical fluctuation in voltage of the battery or due to temperature changes. 
Now we will plot $\sum_{i,j}|t'_{ij}|^2$ as a function of $ \phi/\phi_0 $ for different ${Em^*a^2\over{\hbar}^2} $
 values and we will show that the behavior is also independent of incident energy when 
we employ evanescent modes. This is not the case with propagating modes where changes of 
incident energy result in just as much diversity that we get on changing $ l_1 $ and $ l_2 $
 and hence not shown here. 
\begin{figure}[H]
\centering
\includegraphics[height=6truecm,width=8truecm]{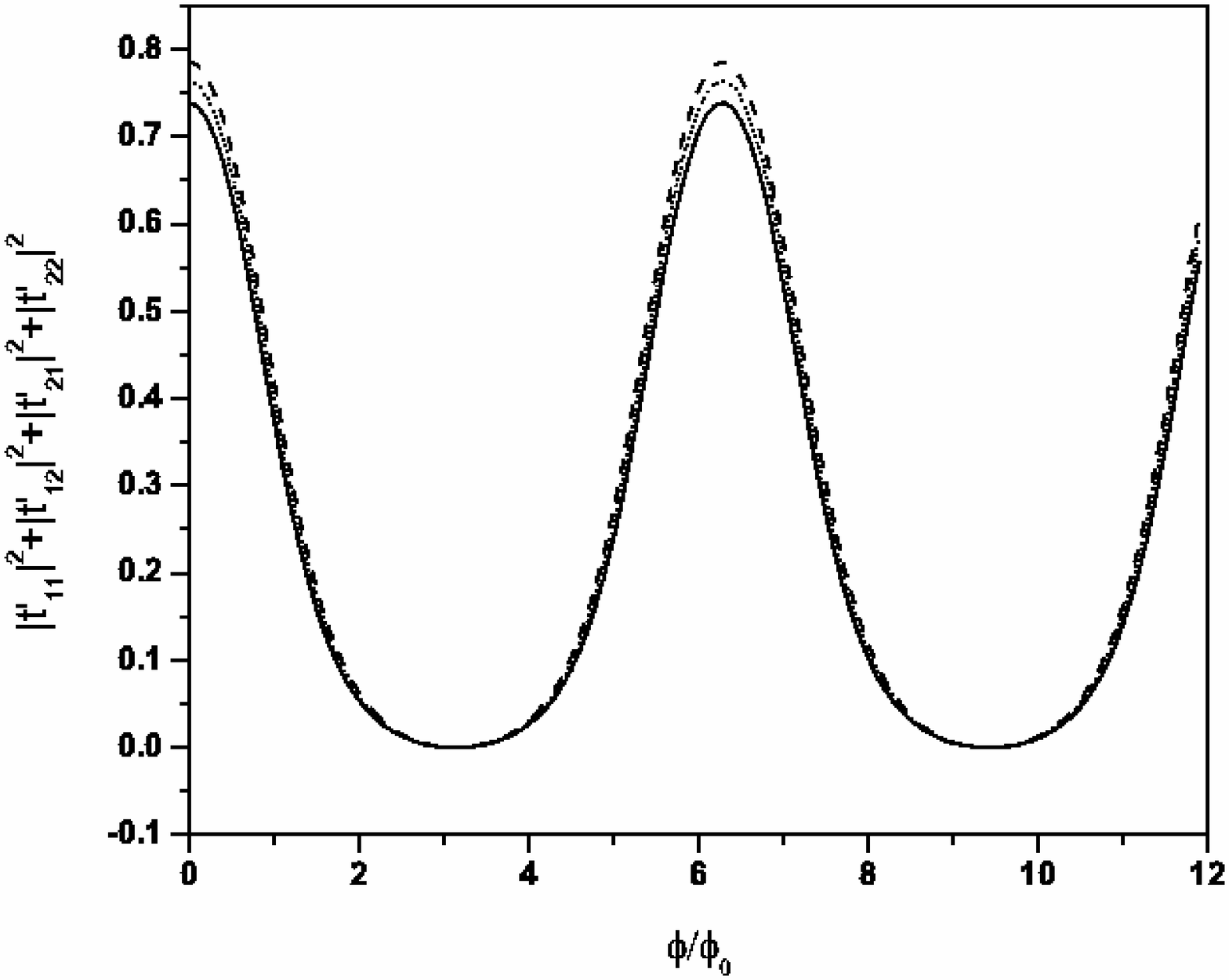}
\caption{\small The figure shows a plot of $\sum_{i,j}|t'_{ij}|^2$ as a function 
of $ \phi/\phi_0 $ for different incoming energy. Here $ \phi_0 = hc/e $. The 
constant potential $ V_0 $ of the ring is such that 
$ V_0 ={40{\hbar}^2\over em^*a^2} $. With this choice $q_1 $ and $q_2$ are 
both imaginary. Thus, we are considering in this case two evanescent modes. 
The solid line is $\sum_{i,j}|t'_{ij}|^2$ for   $ E = {45{\hbar}^2\over m^*a^2} $. 
The dotted line is $\sum_{i,j}|t'_{ij}|^2 $ for   $ E = {46{\hbar}^2\over m^*a^2} $. 
The dashed line is $\sum_{i,j}|t'_{ij}|^2$ for  $ E = {47{\hbar}^2\over m^*a^2} $. 
For all energy values $\sum_{i,j}|t'_{ij}|^2$ as a function of $ \phi/\phi_0 $ have the 
same nature. That means switching action is independent of incident energy.}
\label{fig10}
\end{figure}
In Fig. \ref{fig10}, two values of incident energy and electrostatic potential are so 
chosen that both channels are evanescent. Here again we find that  
$ \sum_{i,j}|t'_{ij}|^2$ as a function of $ \phi/\phi_0 $ is roughly independent 
of incident energy and the drop is almost 75-80\%. Since in the evanescent mode switching 
action is independent of all parameters, switch can become stable, efficient and robust

So far we have considered two propagating, one propagating - one evanescent and two 
evanescent modes separately. For all these cases the total ring length were the same, 
we only changed the relative ratio of arm lengths. In these cases we have shown that when 
there are propagating modes then the peaks are shifting and the depth of the valleys are 
changing from shallow to deep. Resonances are determined by the total ring length. If we 
consider cases where the total ring length does not remain same then one can get even more 
diverse behavior.
\begin{figure}[H]
\centering
\includegraphics[height=6truecm,width=8truecm]{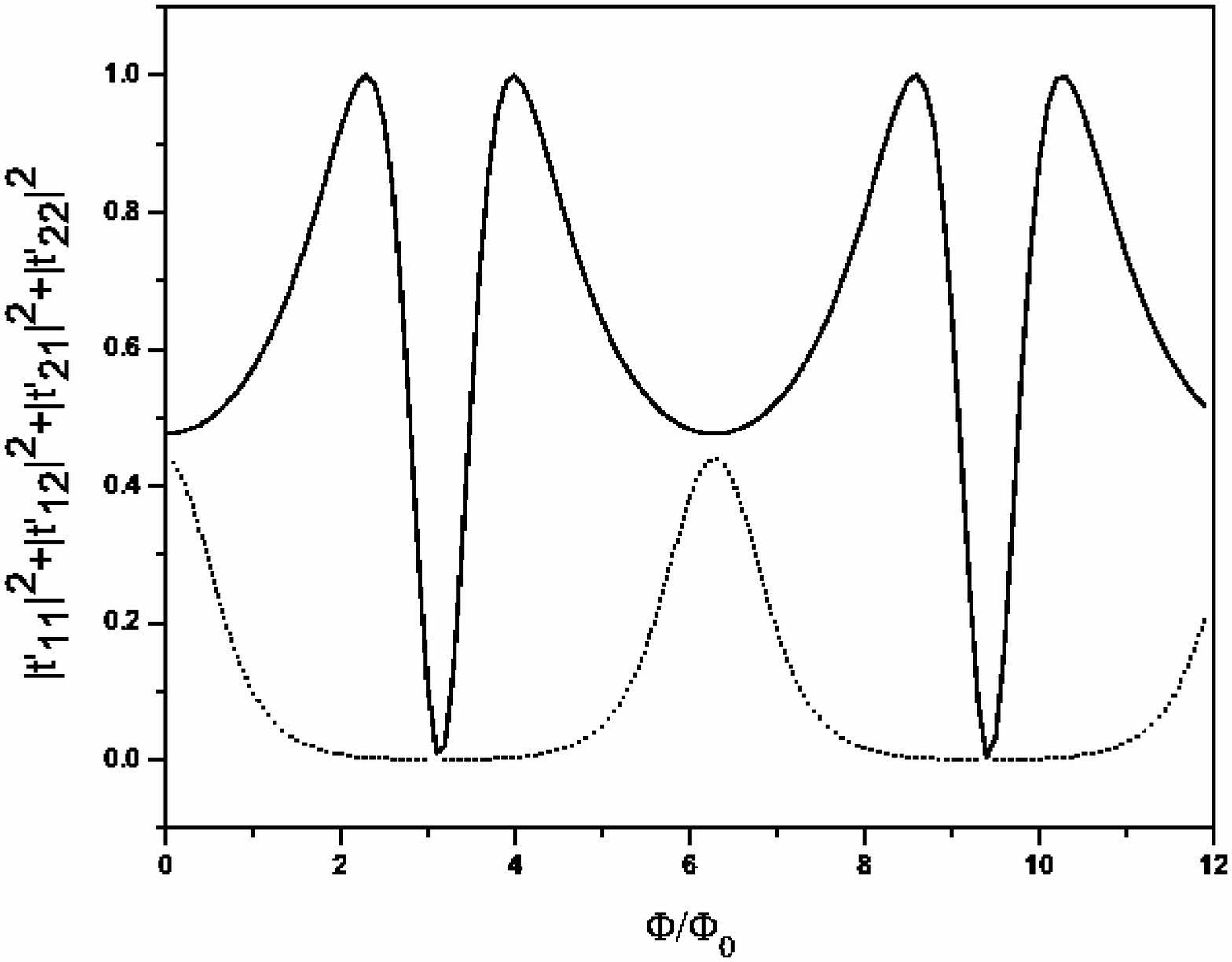}
\caption{\small The figure shows a plot of  $\sum_{i,j}|t'_{ij}|^2$  as a function 
of $ \phi/\phi_0 $ for two choice of total ring lengths. Here $ \phi_0 = hc/e $. 
The incoming electrons have energy  $ E = {45{\hbar}^2\over m^*a^2} $, the constant 
potential $ V_0 $ of the ring is $ 0 $. With this choice $q_1 $ and $q_2$ are both real. 
Thus, we are considering in this case two propagating modes. The solid line is 
$ \sum_{i,j}|t'_{ij}|^2$ for $ L/a =10 $ ($l_1/a=5$, $l_2/a=5$) and the dashed line is 
$\sum_{i,j}|t'_{ij}|^2 $ for $ L/a = 8 $ ($l_1/a=4$, $ l_2/a=4$).}
\label{fig11}
\end{figure}

In Fig. \ref{fig11} we have considered both the channels to be propagating with two choices
 of $ (l_1+l_2) $. We have shown here that resonance position of the solid line is different
 from the resonance position of the dotted line. The solid line has a valley where the dotted
 line has a peak. But again if we use evanescent modes then changes in total ring length
can not result in diverse behavior.
\begin{figure}[H]
\centering
\includegraphics[height=6truecm,width=8truecm]{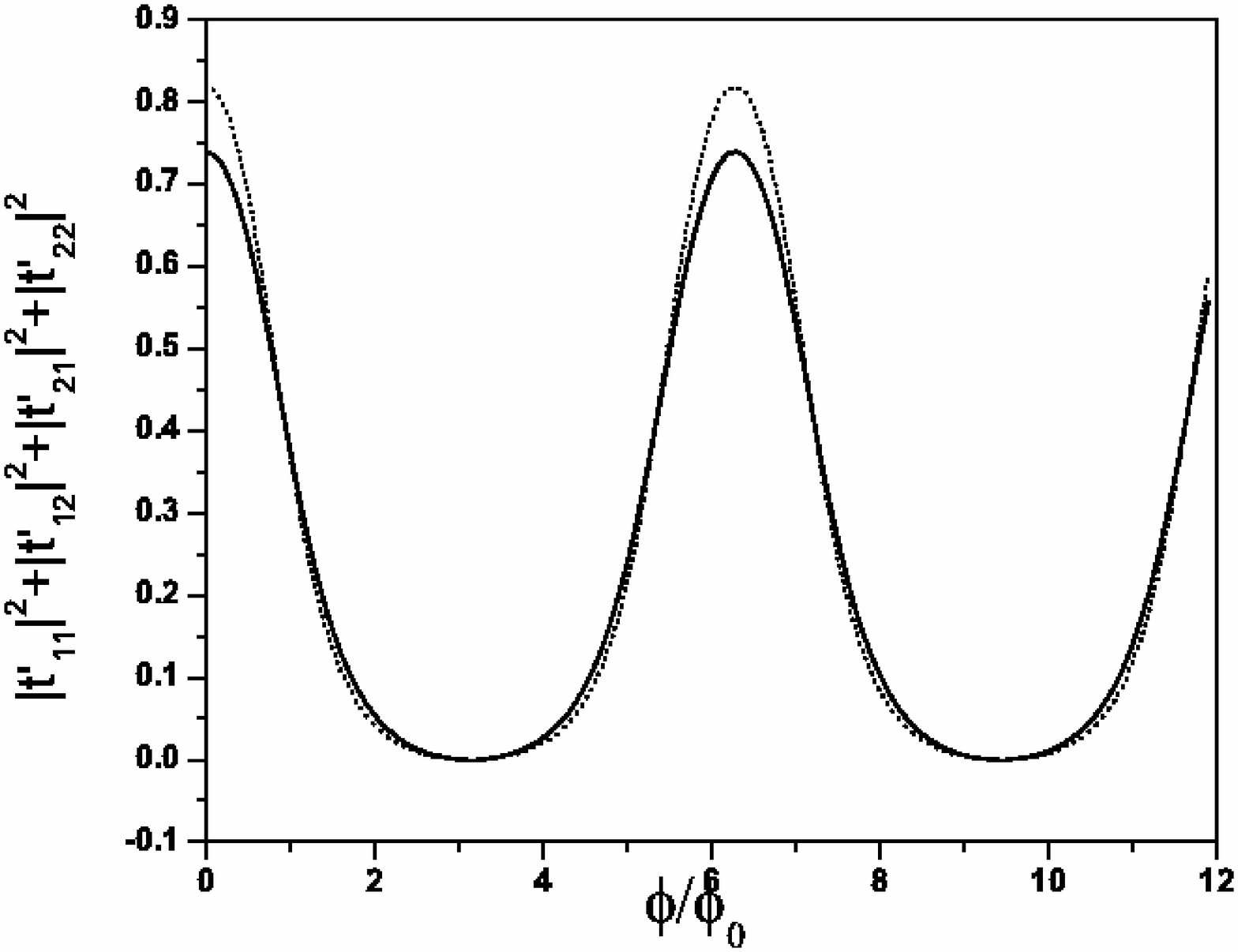}
\caption{\small The figure shows a plot of $\sum_{i,j}|t'_{ij}|^2$ as a 
function of $ \phi/\phi_0 $. Here $ \phi_0 = hc/e $. The incoming electrons
 have energy  $ E = {45{\hbar}^2\over m^*a^2} $, the constant potential 
$ V_0 $ of the ring is such that $ V_0 ={40{\hbar}^2\over em^*a^2} $.
 With this choice $q_1 $ and $q_2$ are both imaginary. 
Thus, we are considering in this case two evanescent modes. 
The solid line is $ \sum_{i,j}|t'_{ij}|^2$ for 
$ L/a =0.2 $ ($l_1/a=0.1$, $l_2/a=0.1$) and the dashed line is
 $\sum_{i,j}|t'_{ij}|^2 $ for $ L/a = 0.16 $ ($l_1/a=0.08$, $ l_2/a=0.08$).}
\label{fig12}
\end{figure}

In Fig. \ref{fig12} we have considered both the channels to be evanescent 
with two choices of $ (l_1+l_2) $. Here the nature of the solid curve and 
the dotted curve are the same as that obtained so far. We have shown in
 previous figures that if we use evanescent modes, conductance does not
 depend on the relative ratios of arm lengths and Fermi energy and in this 
Fig. \ref{fig12} we have shown that the conductance also does not depend on 
the total ring length of the Aharonov - Bohm ring.
\section{ Conclusions}
In this work we have studied two channel (transverse modes) Aharonov - Bohm ring. 
When we consider both the channels to be propagating then we have shown that if 
we change the parameters such as the total ring length, relative ratio of arm lengths, 
Fermi energy etc., the behavior of the conductance becomes diverse in nature 
in different cases. Similar situation arises if we take one propagating and one 
evanescent modes. In presence of channel mixing the modes are not independent. 
Resonance in the propagating channel boost up density of states in the evanescent 
channel and hence the evanescent channel also becomes highly conducting. Such diverse 
behavior supports Landauer's claim that switch action based on interference principle 
are not stable and practical. Finally we have considered both the modes to be evanescent 
along the Aharonov - Bohm ring. Here we have found that conductance is qualitatively as 
well as quantitatively same for all variations in parameters like total ring length, 
relative ratio of arm lengths, Fermi energy etc.  We can obtain appreciable changes in 
conductance when using evanescent modes. Different channels add up coherently 
and so by using larger and larger number of evanescent channels we can enhance the 
percentage drop in conductance and hence efficiency of the switch. Change in impurity 
configuration effectively changes the total ring length and the relative ratio of arm lengths. 
Rise or drop in temperature effectively changes Fermi energy and hence wavelength. 
Therefore, conductance behavior will be same if we change the impurity configuration or 
temperature when we use evanescent modes. Propagation is associated with phase changes 
which do not arise in case of evanescent modes. In evanescent modes, phase changes are 
due to Aharonov - Bohm effect only. Periodicity is always $ \phi_0 $ in absence of other 
competing source of phase changes. Conductance being a symmetric function of flux is 
a function of $ \cos (n\phi/\phi_0) $. Therefore within a period (0 to $2\pi$) conductance 
is maximum at zero flux, then it goes through a deep minimum and rise again to a 
maximum value. We can assign the conductance maximum as on state and conductance minimum as off 
state of a switch signifying 1 and 0 operation in Boolean algebra. Thus we can conclude 
that if we employ evanescent modes only, we may be able to build stable, 
efficient and robust quantum switches.

Earlier works have proposed possibility of switch action with other geometric configurations apart 
from Aharonov - Bohm ring such as T - shaped structure \cite{sols} etc. We may also expect that 
if we employ evanescent modes in other geometries, conductance will be independent of sample parameters. 
This is because propagation along evanescent channels are not associated with phase changes. 
Phase changes can only be induced by external stimuli which is electrostatic potential in case of Ref 5.

\section*{acknowledgment}
One of us (PSD) would like to thank Prof. A. M. Jayannavar for useful discussions.


\begin{thebibliography}{99}
\bibitem{datta} S. Datta, {\it Electronic Transport in Mesoscopic Systems} (Cambridge: Cambridge University Press).
\bibitem{dat85} S. Datta, M. R. Melloch, S. Bandyopadhyay, R. Noren, M. Vaziri, M. Miller and R. Reifenberger. 
{\it Phys. Rev. Lett} {\bf 55}, 2344(1985); S. Datta, M. R. Melloch, S. Bandyopadhyay, and M. S. Lundstrom, 
{\it Appl. Phys. Lett} {\bf 48}, 487 (1986).
\bibitem{dat90} S. Datta and B. Das, {\it Appl. Phys. Lett} {\bf 56}, 665 (1990).
\bibitem{datrpp} S. Datta and M. J. McLennan, {\it Rep. Prog. Phys.} {\bf 53}, 1003(1990).
\bibitem{sols} F. Sols, M. Macucci, V. Ravoili, and K. Hess, {\it Appl. Phys. Lett.}, {\bf 54}, 350 (1990); {\it J. Appl. Phys.} {\bf 66}, 3892 (1989).
\bibitem{sub} S. Subramaniam, S. Bandopadhyay, and W. Porod, {\it J. Appl. Phys.} {\bf 68}, 4861(1990).
\bibitem{land} R. Landauer, {\it Physics Today} {\bf 42}, 10, 119(1989).
\bibitem{bcg} B. C. Gupta, P. Singha Deo and A. M. Jayannavar, {\it Int. J. Mod. Phys. B} {\bf 10} 3595(1996).
\bibitem{deo} A. M. Jayannavar and P. Singha Deo, {\it Mod. Phys. Lett B} {\bf 8}, 301(1994).
\bibitem{mol} B. Moln$\acute{a}$r, F. M. Peeters and P. Vasilopoulos, {\it Phys. Rev B} {\bf 69}, 155335(2004). 
\bibitem{pet} B. Moln$\acute{a}$r, P. Vasilopoulos and F. M. Peeters, {\it Appl. Phys. Lett} {\bf 85}, 4 (2004).
\bibitem{pet1} P. F$\ddot{o}$ldi, B. Moln$\acute{a}$r, M. G. Benedict and F. M. Peeters, {\it Phys. Rev B} {\bf 71}, 033309(2005). 
\bibitem{pet2} B. Moln$\acute{a}$r, P. Vasilopoulos and F. M. Peeters, {\it Phys. Rev B} {\bf 72}, 075330(2005).
\bibitem{pet3} P. F$\ddot{o}$ldi, O. K$\acute{a}$lm$\acute{a}$n, M. G. Benedict, and F. M. Peeters, {\it Phys. Rev B} {\bf 73}, 155325(2006).
\bibitem{vasi} P. Vasilopoulos, O. K$\acute{a}$lm$\acute{a}$n, F. M. Peeters, and M. G. Benedict, {\it Phys. Rev B} {\bf 75}, 035304(2007).
\bibitem{kal} O. K$\acute{a}$lm$\acute{a}$n, P. F$\ddot{o}$ldi, M. G. Benedict, F. M. Peeters, {\it Physica E} {\bf 40}, 567(2008).
\bibitem{jos} S. K. Joshi, D. Sahoo and A. M. Jayannavar, {\it Phys. Rev B} {\bf 64}, 075320(2001).
\bibitem{ben2} C. Benjamin and A. M. Jayannavar, {\it Int. J. Mod. Phys. B} {\bf 16}, 1787(2002).
\bibitem{she} S. Sengupta Chowdhury, P. Singha Deo, A. K. Roy and M. Manninen,{\it New Journal of Physics}  {\bf 10}, 083014(2008).
\bibitem{che} H. F. Cheung, Y. Gefen, E. K. Riedel, and W. H. Shih, {\it Phys. Rev B} {\bf 37}, 6050(1988).
\bibitem{wee} B. J. van Wees, H. van Houten, C. W. J. Beenakker, J. G. Williamson, L. P. Kouwenhoven, D. van der Marel, and C. T. Foxon, {\it Phys. Rev. Lett.} {\bf 60}, 848(1988).
\bibitem{wha} D. A. Wharam, T. J. Thornton, R. Newbury, M. Pepper, H. Ahmed, J. E. Frost, D. G. Hasko, D. C. Peacock, D. A. Ritchie, and G. A. C. Jones, {\it J. Phys. C} {\bf 21} L209(1988).
\bibitem{but} M. B$\ddot{u}$ttiker, {\it Phys. Rev. B}, {\bf 32}, 1846(1985).
\bibitem{pin} M. B$\ddot{u}$ttiker, Y. Imry, R. Landauer, and S. Pinhas  {\it Phys. Rev. B} {\bf 311}, 6207(1985).
\bibitem{merz} E. Merzbacher, {\it Quantum Mechanics}, 3rd ed. (Wiley, New York, 1997). 
\bibitem{pas} H. M. Pastawski, A. Rojo and C. Balseiro, {\it Phys. Rev B} {\bf 37}, 6246(1988).
\end{thebibliography}
\end{document}